# Study on the fiber-bundle topological properties of band structure of eigenvalue under Periodic boundary conditions

Xiangxin Meng[1,†] and Yurui Fang[1,*]

[1] *School of Physic, Dalian University of Technology.*

[†] *xxmeng@mail.dlut.edu.cn*

**Corresponding authors: yrfang@dlut.edu.cn*

**Abstract**

Noether's theorem establishes the fundamental relation between continuous symmetries and associated conserved charges, while periodic translational symmetry provides the physical foundation for Bloch's theorem. In this work, we investigate the eigenvalue structures arising from fiber bundles and systematically classify their topological properties using group theory. Periodic Hermitian operators naturally generate such dispersions defined over the Brillouin zone. At high-symmetry points, the total dimension of the eigenvalue branches remains invariant. We classify the fiber-bundle subspaces at band crossings to distinguish genuinely intersecting and locally degenerate fiber eigenvalues, thereby uncovering hidden additional symmetries. Furthermore, topological invariants such as Chern numbers can be rigorously defined for individual fiber eigenvalues, independently of band intersections. This classification scheme applies directly to wave band structures in periodic potentials and may be extended to photonic crystals, higher-dimensional systems, and particle physics.

## Introduction

Noether's theorem states that every continuous symmetry of a physical system corresponds to a conserved quantity. Common examples are energy conservation associated with continuous time-translation symmetry and momentum conservation associated with continuous spatial-translation symmetry.

Meanwhile, according to Bloch's theorem, a periodic potential in space gives rise to a momentum jump in the phase of the wave function, which produces a band-gap structure in energy—i.e., the band structure of a photonic crystal. Likewise, a periodic potential in time produces an energy jump in the phase of the wave function, which produces a band-gap structure in momentum—i.e., a time photonic crystal. It is then natural to deduce that, for any continuous quasiparticle motion or field wave, a periodic potential breaks a symmetric current, giving rise to a charge jump that induces a phase jump of the wave function and thus forms a band-gap structure in the dispersion relation. On this basis, the band structure can be further classified.

Here, we intend to demonstrate the above idea and to classify the topological properties of the band structure using group theory.

## Demonstration

We start form two important theorems that will be used in the following discussion:

**Spectral theorem.** Let $\mathcal{H}$ be a Hilbert space and let $\boldsymbol{A}$ be a self-adjoint operator in $\mathcal{B}(\mathcal{H})$. Then there exists a unique spectral measure[1,2]

$$P_A: \mathfrak{B}(\sigma(\mathbf{A})) \to \mathcal{P}(\mathrm{H})$$

so that

$$\boldsymbol{A} = \int_{\sigma(\boldsymbol{A})} \lambda dP_{\boldsymbol{A}}(\lambda) \tag{1}$$

Meantime, for every bounded Borel function $f$, $f(\mathbf{A}) = \int \mathrm{f}(\lambda)\mathrm{d}\mathbf{E}(\lambda)$, where the integration region is $\sigma(\mathbf{A})$.

**Riesz projection theorem.** let $\boldsymbol{B}(\lambda)$ be a family of self-adjoint projections depending on the parameter $\lambda$, and let $b_0$ be an isolated spectral point of $\boldsymbol{B}(\lambda_0)$. Take a positively oriented simple closed $\Gamma$ that encloses $b_0$ and contains no other spectrum of $\boldsymbol{B}(\lambda_0)$, such that $\Gamma$ lies entirely in the resolvent operator is continuous in the operator norm:

$$\left\| \left(z - \boldsymbol{B}(\lambda)\right)^{-1} - \left(z - \boldsymbol{B}(\lambda_0)\right)^{-1} \right\| \to 0$$

then:

$$\boldsymbol{P}(\lambda) = \frac{1}{2\pi i} \oint \left(z - \boldsymbol{B}(\lambda)\right)^{-1} dz \tag{2}$$

**1. Continuity of Fibers and Fiber Eigenvalues**

For self-adjoint operators **A** and **B** defined on a Hilbert space $\mathcal{H}$, the spectral theorem guarantees the existence of a projection-valued measure $P_{\boldsymbol{A}}(\lambda)$ for **A**, so that **A** can be represented as

$$\boldsymbol{A} = \int_{\sigma(\boldsymbol{A})} \lambda dP_{\boldsymbol{A}}(\lambda) \tag{3}$$

The space $\mathcal{H}$ can then be decomposed as a direct integral; write $\mathcal{H} = \int_{\sigma(\boldsymbol{A})}^{\oplus} \mathcal{H}_\lambda d\mu(\lambda)$. On the fiber $\mathcal{H}_\lambda$, the operator **A** can be further represented as[3, 4]

$$A = \int_{\sigma(\boldsymbol{A})}^{\oplus} \lambda I_\lambda d\mu(\lambda) \tag{4}$$

where $\sigma(\boldsymbol{A})$ is the spectrum of **A**, $\mathcal{H}_\lambda$ is the fiber space, and $\mu$ is the spectral measure. If $[\mathbf{A}, \mathbf{B}] = 0$, then **B** does not mix different spectral fibers of **A**; it acts fiberwise in the direct-integral decomposition of **A**. Its form is as follows

$$\boldsymbol{B} = \int_{\sigma(\boldsymbol{A})}^{\oplus} \boldsymbol{B}(\lambda) d\mu(\lambda) \tag{5}$$

where $\boldsymbol{B}(\lambda)$ is an operator on the fiber $\mathcal{H}_\lambda$.

The concrete construction of $\mathbf{B}(\lambda)$ is as follows: first find the operator $\boldsymbol{U}$ that diagonalizes $\boldsymbol{A}$, then restrict it to the fiber; in this case $\mathbf{B}(\lambda) = \boldsymbol{U}\boldsymbol{B}\boldsymbol{U}^{-1}|_{\mathcal{H}_\lambda}$

For convenience, we impose the additional assumption that on each fiber $\mathcal{H}_\lambda$, the operator $\boldsymbol{B}(\lambda)$ has compact resolvent. Then the eigenvalues of $\boldsymbol{B}(\lambda)$ are discrete, each eigenvalue has finite multiplicity, and there are no finite accumulation points.[5 3]

Arrange the eigenvalues of $\boldsymbol{B}(\lambda)$ in decreasing order: (Here it is assumed that $\boldsymbol{B}(\lambda)$ is bounded below; all operators below are assumed to satisfy this condition.)

$$0 < E_1(\lambda) < E_2(\lambda) < \cdots \tag{6}$$

where $\mathrm{E_i}(\lambda)$ denotes the eigenvalue, and we write $\mathrm{V_i}(\lambda) = \ker(\mathbf{B}(\lambda) - \mathrm{E_i}(\lambda)\mathbf{I})$.

In following part, we will prove that the fiber eigenvalues are continuous under the given conditions. To this end, we explain the lemmas and concepts to be used.

**Lemma 1.** For any $x \in \mathcal{H}$ and $\boldsymbol{T} \in L(\mathcal{H})$, the following inequality holds: $\|\boldsymbol{T}x\| \leq \|x\|\|x\|$.

**Definition.** The Rayleigh quotient of an operator **T** is $\frac{\langle Tx,x\rangle}{\langle x,x\rangle}$ [6].

**Lemma 2 (Courant-Fischer min-max theorem).** For any $x \in \mathcal{H}$ and $\boldsymbol{T} \in L(\mathcal{H})$, assuming $\boldsymbol{T}$ has compact resolvent, its kth eigenvalue is[2, 6]

$$\alpha_k = \min_{\substack{S \subset Q(A) \\ \dim S = k}} \max_{0 \neq x \in S} \frac{\langle \boldsymbol{T}x, x\rangle}{\langle x, x\rangle} \tag{7}$$

The region $Q(\boldsymbol{A})$ is the domain in which the Rayleigh quotient of T is defined.

For $x \in \mathcal{H}$, let its projection onto $\mathcal{H}_\lambda$ be $\boldsymbol{P}(\lambda)x \in \mathcal{H}_\lambda$; then $x(\lambda)$ may be regarded as a section of the fiber. If $\xi \in \mathcal{H}_{\lambda_0}$, define the section $\tilde{\xi} = \begin{cases} \xi & \lambda = \lambda_0 \\ 0 & \lambda \neq \lambda_0 \end{cases}$. Then $B(\lambda_0)$ is naturally extended to $\mathcal{H}$; denote the extended operator by $\widetilde{\boldsymbol{B}(\lambda)}$. Its action on sections is naturally defined as $\widetilde{\boldsymbol{B}(\lambda_0)}x(\lambda) = \boldsymbol{B}(\lambda_0)\delta(\lambda - \lambda_0)x(\lambda)$. Obviously, $\left\|\widetilde{\boldsymbol{B}(\lambda)}\right\|_{\mathcal{H}} = \|\boldsymbol{B}(\lambda)\|_{\mathcal{H}_\lambda}$.

**Theorem 1.** If $\widetilde{\boldsymbol{B}(\lambda)}$ is continuous with respect to $\lambda$, then $E_i(\lambda)$ $i \in \mathbb{N}$ is continuous as a function of $\lambda$ for all i in $N$.

**Proof.** From Lemmas 1 and 2, we obtain

$$\left| E_i(\lambda) = \min_{\substack{S \subset Q(A) \\ \dim S = i}} \max_{0 \neq x \in S} \frac{\langle \widetilde{B(\lambda)} x, x \rangle}{\langle x, x \rangle} \right| < \left\| \widetilde{\boldsymbol{B}(\lambda)} \right\| \tag{8}$$

Therefore, $|E_i(\lambda) - E_i(\lambda')| < \left\| \widetilde{\boldsymbol{B}(\lambda)} - \widetilde{\boldsymbol{B}(\lambda')} \right\|$. So for a Hermite operator, which is self-adjoint, gives a continuous eigenvalue dispersion relation. If the operator is periodic, the eigenvalue dispersion relation will be continuous in Brillion zone.

**2. Properties of Fiber-Bundle Subspaces**

Here we will introduce two important properties of the eigenvalue fiber bundle, which provide essential support for defining a topological structure on it.

**Lemma 3.** If projection operators $\boldsymbol{P}$ and $\boldsymbol{Q}$ satisfy $\|\boldsymbol{P} - \boldsymbol{Q}\| < 1$, then they have the same rank.

**Proof.** For any $x \in ran\boldsymbol{Q}$, we have $\boldsymbol{Q}x = x$, hence $\|\boldsymbol{P}x - x\| = \|\boldsymbol{P}x - \boldsymbol{Q}x\| \leq \|\boldsymbol{P} - \boldsymbol{Q}\|\|x\| < \|x\|$. Therefore, $\boldsymbol{P}x \neq x, x \neq 0$, which shows that $\boldsymbol{P}|_{ranQ}: ran\boldsymbol{Q} \to ran\boldsymbol{P}$ is injective,

$$dim\, ran\boldsymbol{Q} \leq dim\, ran\boldsymbol{P}$$

At the same time, interchanging $\boldsymbol{P}$ and $\boldsymbol{Q}$ gives $dim\, ran\boldsymbol{P} \leq dim\, ran\boldsymbol{Q}$. Hence, $dim\, ran\boldsymbol{Q} = dim\, ran\boldsymbol{P}$.

**Theorem 2.** By the Riesz representation theorem, one can choose a contour $\Gamma$ surrounding the eigenvalues and define

$$\widetilde{\boldsymbol{P}(\lambda)} = \frac{1}{2\pi i} \oint \left(z - \widetilde{\boldsymbol{B}(\lambda)}\right)^{-1} dz \tag{9}$$

If the spectral gap is preserved and $\boldsymbol{B}(\lambda)$ is continuous, then: $\widetilde{\boldsymbol{P}(\lambda)}$ is a continuous projection in $\mathcal{H}$; the rank of a continuous projection is locally constant; and when $\sigma(\boldsymbol{A})$ is a connected interval, $dimV_i(\lambda) =$ constant.

**Proof.** Since $\widetilde{\boldsymbol{B}(\lambda)}$ is a self-adjoint operator, there exists a contour $\Gamma$ that separates a group of eigenvalues $\sigma_\Gamma(\lambda)$ from the rest of the spectrum.

$\widetilde{\boldsymbol{B}(\lambda)}$ is continuous in the norm-resolvent sense: $\left\| \left(z - \widetilde{\boldsymbol{B}(\lambda)}\right)^{-1} - \left(z - \widetilde{\boldsymbol{B}(\lambda_0)}\right) \right\| \to 0$. Since

$\Gamma$ is compact, this holds uniformly for all $z \in \Gamma$. Define $\widetilde{\boldsymbol{P}(\lambda)} = \frac{1}{2\pi i}\oint\left(z - \widetilde{\boldsymbol{B}(\lambda)}\right)^{-1} dz$,

$$\widetilde{\boldsymbol{P}(\lambda)} - \widetilde{\boldsymbol{P}(\lambda_0)} = \frac{1}{2\pi i}\oint\left(z - \widetilde{\boldsymbol{B}(\lambda)}\right)^{-1} - \left(z - \widetilde{\boldsymbol{B}(\lambda_0)}\right)^{-1} dz \tag{10}$$

Clearly, $\widetilde{\boldsymbol{P}(\lambda)}$ is a norm-continuous operator. By the spectral theorem, we know $\widetilde{\boldsymbol{B}(\lambda)} = \int \mu d\boldsymbol{E}(\mu)$. The resolvent can be written as $\left(z - \widetilde{\boldsymbol{B}(\lambda)}\right)^{-1} = \int \frac{1}{z-\mu} d\boldsymbol{E}(\mu)$; substituting this into $\widetilde{\boldsymbol{P}(\lambda)}$ gives $\widetilde{\boldsymbol{P}(\lambda)} = \frac{1}{2\pi i}\oint \int \frac{1}{z-\mu} d\boldsymbol{E}(\mu)\, dz$.

Interchange the order of integration (assuming that this is justified).

$$\widetilde{\boldsymbol{P}(\lambda)} = \frac{1}{2\pi i}\int \oint \frac{1}{z-\mu} dz dE(\mu) \tag{11}$$

By the Cauchy integral formula,

$$\frac{1}{2\pi i}\oint \frac{1}{z-\mu}\, dz = \begin{cases} 1 & ,\mu \text{ inside } \Gamma \\ 0 & ,\mu \text{ outside } \Gamma \end{cases} \tag{12}$$

$\boldsymbol{P}(\lambda)$ is the spectral projection onto the interior of $\Gamma$, and is naturally an orthogonal projection:

$$\widetilde{\boldsymbol{P}(\lambda)}\mathcal{H} = \bigoplus_{\mu \in \sigma_\Gamma(\lambda)} ker\left(\widetilde{\boldsymbol{B}(\lambda)} - \mu \boldsymbol{I}\right) \tag{13}$$

Since the interior of $\Gamma$ contains only one isolated eigenvalue $E_i(\lambda)$, we have

$$\boldsymbol{P}(\lambda)\mathcal{H} = ker\left(\widetilde{\boldsymbol{B}(\lambda)} - E_i(\lambda)\boldsymbol{I}\right) = V_i(\lambda) \tag{14}$$

From the lemma we know that $\boldsymbol{P}(\lambda)$ is locally constant. If the interval is connected, a locally constant function on a connected interval must be constant on the whole interval. Therefore, $ker\left(\widetilde{\boldsymbol{B}(\lambda)} - E_i(\lambda)\boldsymbol{I}\right) = \text{constant}$ holds for every $\lambda$ in the interval.

**Theorem 3.** Suppose the fiber eigenvalues $E_i(\lambda)$ and $E_j(\lambda)$ intersect at $\lambda_0$ and are isolated at all other values of $\lambda$. Then the sub-eigenvalue at this point is $E_i(\lambda_0) = E_j(\lambda_0) = E_0$, and the dimension of the corresponding subspace $V(\lambda_0)$ equals the sum of the dimensions of the branches.[7] [8]

**Proof.** Choose a contour $\Gamma$ that contains exactly the two eigenvalue branches and no others; this is possible for operators with compact resolvent.

$$\boldsymbol{P}(\lambda) = \frac{1}{2\pi i}\oint\left(z - \widetilde{\boldsymbol{B}(\lambda)}\right)^{-1} dz \tag{15}$$

For $\lambda \neq \boldsymbol{\lambda_0}$, $\Gamma$ contains two distinct eigenvalues $E_i(\lambda)$ and $E_j(\lambda)$, so

$$P(\lambda)\mathcal{H} = V_i(\lambda) \oplus V_j(\lambda) \tag{16}$$

Therefore, $rank\boldsymbol{P}(\lambda) = dimV_i(\lambda) + dimV_j(\lambda)$. When $\lambda = \lambda_0$, $\boldsymbol{P}(\lambda_0)\mathcal{H} = ker\big(\widetilde{\boldsymbol{B}(\lambda_0)} - E_0\boldsymbol{I}\big)$. Since $\boldsymbol{P}(\lambda)$ is a continuous family of projections, it is locally constant. As $\lambda \longrightarrow \lambda_0$, $rank\boldsymbol{P}(\lambda) = rank\boldsymbol{P}(\lambda_0)$. Therefore, $dim\,V(\lambda_0) = dimV_i(\lambda) + dimV_j(\lambda)$

From the above, the fiber eigenvalues are continuous. For an isolated single fiber eigenvalue, the dimension of the corresponding subspace is fixed. For crossing or locally coincident fiber eigenvalues, the dimension of the corresponding subspace jumps. However, at such points the subspace corresponding to all eigenvalues is the sum of the subspaces corresponding to the crossing or coincident fiber eigenvalues.

Thus for a band structure without gap, like at a Dirac point, we prove that the total dimension of different branches of eigenvalues keeps the same, and therefore we can further distinguish the fiber-bundle subspaces at the overlapped point in the band structure.

### 3. Marking Spaces by Group Representations

Now we introduce additional information that allows us to distinguish further the subspaces corresponding to crossing or locally coincident fiber eigenvalues.

Assume that all symmetry information of the operator is known. Let the symmetry group be $\mathfrak{g}$, with group elements $g \in \mathfrak{g}$; the corresponding operator on the space is $\boldsymbol{U}_g$. Then

$$[\boldsymbol{U}_g, \boldsymbol{B}] = 0 \tag{17}$$

$\boldsymbol{B} = \boldsymbol{U}_g\,\boldsymbol{B}\boldsymbol{U}_g{}^{-\boldsymbol{1}}$. For the spectrum, $\sigma(\boldsymbol{B}) = \sigma\big(\boldsymbol{U}_g\,\boldsymbol{B}\boldsymbol{U}_g{}^{-\boldsymbol{1}}\big)$, which shows that the eigenvalue spectrum is invariant.

In the following discussion we always assume that $\mathfrak{g}$ is a finite or compact group; this is essentially the case for the groups encountered in physics. With the group information, the elements of $\mathfrak{g}$ can be classified into two parts. If an element $\mathfrak{g}$ of $g$ satisfies $\boldsymbol{U}_g\,\boldsymbol{A}\boldsymbol{U}_g{}^{-\boldsymbol{1}} = f_g(\boldsymbol{A})$, then $g$ is said to be covariant with respect to **B**. All covariant elements of $\mathfrak{g}$ form a subgroup denoted $\mathfrak{g}'$, called the covariant subgroup of **B**. The subgroup formed by the non-covariant elements with respect to **B** is called the non-covariant subgroup. We will illustrate the reason of such classification with the following theorem.

**Theorem 4.** The operators corresponding to the elements $g'$ in $g$ satisfy $\boldsymbol{U}_g\mathcal{H}_\lambda = \mathcal{H}_{\lambda'}$ for all $\lambda$ (strictly speaking, for almost every $\lambda$; however, for a real physical system, one may regard it as holding for every $\lambda$).

**Proof.** Construct the operator $\boldsymbol{A'} = \boldsymbol{U}_g\,\boldsymbol{A}\boldsymbol{U}_g{}^{-\mathbf{1}}$. By the spectral theorem, $\boldsymbol{P}_{\boldsymbol{A'}}(\boldsymbol{\Delta}) = \boldsymbol{U}_g\,\boldsymbol{P}_{\boldsymbol{A}}(\boldsymbol{\Delta})\boldsymbol{U}_g{}^{-\mathbf{1}}$, where $\Delta$ is the Borel set generated by open sets in $\sigma(\boldsymbol{A})$. Since $\boldsymbol{A'}$ and $\boldsymbol{A}$ are unitarily equivalent, $\sigma(\boldsymbol{A}) = \sigma(\boldsymbol{A'})$, so

$$\sigma(\boldsymbol{A}) = \sigma\left(f_g(\boldsymbol{A})\right) = f_g\big(\sigma(\boldsymbol{A})\big) \tag{18}$$

Since $\boldsymbol{U}_g\,\boldsymbol{A}\boldsymbol{U}_g{}^{-\mathbf{1}} = f_g(\boldsymbol{A})$, we have $P_{f(\boldsymbol{A})}(\Delta) = P_{\boldsymbol{A}}\left(f_g^{-1}(\Delta)\right)$, $f_g^{-1}(\Delta) = \{\lambda | f_g(\lambda) \in \Delta\}$. Combining the above equalities, $U_g\,P_{\boldsymbol{A}}(\Delta)U_g{}^{-1} = P_{\boldsymbol{A}}\left(f_g^{-1}(\Delta)\right)$; setting $g = f_g^{-1}$, we obtain $U_g\,P_{\boldsymbol{A}}(\Delta)U_g{}^{-1} = P_{\boldsymbol{A}}\big(g(\Delta)\big)$.

From Theorem 4, the relation between $\lambda'$ and $\lambda$ depends on $g$. This allows us to define the map $g\colon \lambda \to \lambda'$. In what follows we denote $\lambda'$ by $g\lambda$. We can then define the action of a group element $g$ on $\sigma(\boldsymbol{A})$, and define the symmetry at a point $\lambda \in \sigma(\boldsymbol{A})$. If $g\lambda = \lambda$, then $\lambda$ is said to satisfy $g$-symmetry.

At the point $\lambda$, the symmetries form a group denoted $G_\lambda$. One can analyze whether the symmetry increases or decreases as $\lambda$ varies (this is meaningful for the variation of eigenvalues). When $\lambda_1$ reaches $\lambda_2$ along a path $\xi \in \sigma(\boldsymbol{A})$, one can define $G_\xi = \{\cap G_\lambda | \lambda \in \xi\}$. This definition could be supplemented with additional mathematical structure; it is unknown whether there is corresponding physics supporting it. The condition for the existence of a fiber eigenvalue connecting $E(\lambda_1)$ and $E(\lambda_2)$ is very likely related to this. Thus here we give a possible conclusion: for any two eigenvalues $E(\lambda_0)$ and $E(\lambda_1)$, and for any path $\xi$ connecting the two points, if the subspaces corresponding to $E(\lambda_0)$ and $E(\lambda_1)$ can both be decomposed into irreducible representations of $G_\xi$, then the condition for a fiber to pass through both points is that, for every path, the irreducible representation types of the two subspaces under $G_\xi$ are the same.

Defining parameter symmetry in this way is more general. Physically, the band structure satisfies all spatial point-group symmetries of the crystal, since extending point-group symmetry to the Brillouin zone is straightforward. However, many abstract symmetries can also be defined on the Brillouin zone in this manner. For instance, time-reversal symmetry can be represented by a sign reversal of the wave vector K, and defining it this way is quite natural. In this way we can further find more symmetry in the Brillion zone in the band structures.

Another important purpose of introducing group symmetry is that it provides an important tool, namely group representation theory. Using group representation theory, the eigenspace of $\boldsymbol{B}$ can be decomposed according to the irreducible representations of the group: $V_i(\lambda) = \oplus\big(M_\alpha(\lambda)\otimes V_{i,\alpha}\big)$, where $M_\alpha(\lambda)$ is the multiplicity space and $\boldsymbol{V}_{i,\alpha}$ is the irreducible representation space. [9-12]

Classify the elements of $\mathfrak{g}'$; let all $\boldsymbol{U}_g$ that commute with **A** form a set. Clearly, the group corresponding to this set is a subgroup of the original group. Denote this set by $\mathfrak{a}$. Then from the above we have

$$\boldsymbol{U}_g = \boldsymbol{B} = \int_{\sigma(\boldsymbol{A})}^{\oplus} \boldsymbol{U}_g(\lambda) d\mu(\lambda) \tag{19}$$

Write $\boldsymbol{U}_g(\lambda) = \boldsymbol{U}_g|_{\mathcal{H}_\lambda}$; then $\big[\boldsymbol{U}_g(\lambda), \boldsymbol{B}(\lambda)\big] = 0$ holds fiberwise. The irreducible representations of this small group also hold fiberwise. Using the branching rule, $V_{i,\alpha} = \oplus\big(M_{\alpha'}(\lambda)\otimes V_{i,\alpha'}\big)$, so the space can now be represented as $V_i(\lambda) = \oplus\big(M_{\alpha'}(\lambda)\otimes V_{i,\alpha'}\big)$. Theorem 5 below will show that, under certain conditions, the irreducible representation $V_{i,\alpha'}$ remains unchanged as the fiber eigenvalue varies. During the variation of the fiber eigenvalue, the multiplicity space $M_{\alpha'}(\lambda)$ varies with it. For convenience, in the following theorem it suffices to prove that the multiplicity of the representation space does not change. Let $m_{\alpha'}(\lambda)$ denote the multiplicity of $V_{i,\alpha'}$.

**Theorem 5.** Suppose $E_i(\lambda)$ is an isolated fiber eigenvalue. Let $\boldsymbol{P}_i(\lambda)$ be the projection operator onto $V_i(\lambda)$, and suppose $U_g(\lambda)$ is norm-continuous with respect to $\lambda$, with $\big[\boldsymbol{U}_g(\lambda), \boldsymbol{B}(\lambda)\big] = 0$, then $m_{\alpha'}(\lambda)$ is locally constant.

**Proof.** Define $\boldsymbol{Q}_{\alpha'}(\lambda) = d_{\alpha'} \int \overline{\chi_{\alpha'}(g)}\, \boldsymbol{U}_g(\lambda) dg$, where $d_{\alpha'}$ is the dimension of the irreducible representation and $\chi_{\alpha'}(g)$ is its character. By definition, $\boldsymbol{Q}_{\alpha'}(\lambda)$ is a projection on the fiber $\mathcal{H}_\lambda$.

Since $\big[\boldsymbol{U}_g(\lambda), \boldsymbol{B}(\lambda)\big] = 0$, as in the proof of the theorem, choose a contour Gamma; by the Riesz projection, $\boldsymbol{P}_i(\lambda)\frac{1}{2\pi i}\oint\big(z - \boldsymbol{B}(\lambda)\big)^{-1} dz$, so

$$U_g(\lambda)\big(z - \boldsymbol{B}(\lambda)\big) = \big(z - \boldsymbol{B}(\lambda)\big)U_g(\lambda) \tag{20}$$

Taking inverses on both sides gives $\left(z-\boldsymbol{B}(\lambda)\right)^{-1}\boldsymbol{U}_g(\lambda)^{-1}=\boldsymbol{U}_g(\lambda)^{-1}\left(z-\boldsymbol{B}(\lambda)\right)^{-1}$; multiplying both sides by $\boldsymbol{U}_g(\lambda)$

$$\left[\boldsymbol{U}_g(\lambda),\left(z-\boldsymbol{B}(\lambda)\right)^{-1}\right]=0 \tag{21}$$

Thus, $[\boldsymbol{Q}_{\alpha'}(\lambda),\boldsymbol{P}_i(\lambda)]=0$. Define $\boldsymbol{P}_{i,\alpha'}(\lambda)=\boldsymbol{P}_i(\lambda)Q_{\alpha'}(\lambda)$. Then $\boldsymbol{P}_{i,\alpha'}(\lambda)$ is again a projection operator, and its image ran $\boldsymbol{P}_{i,\alpha'}(\lambda)$ is the intersection of $V_i(\lambda)$ with the component of type $\alpha'$. Clearly,

$$ran\boldsymbol{P}_{i,\alpha'}(\lambda)=m_{\alpha'}(\lambda)d_{\alpha'} \tag{22}$$

Since $\boldsymbol{P}_{i,\alpha'}(\lambda)$ is a continuous family of projections, the lemma above shows that the rank of $\boldsymbol{P}_{i,\alpha'}(\lambda)$ is locally constant. Consequently, $m_{\alpha'}(\lambda)$ is also locally constant. On the compact set $\sigma(\boldsymbol{A})$, $m_{\alpha'}(\lambda)$ is constant as well.

**Theorem 6.** Let ɑ be the symmetry group acting on the fiber, and suppose the corresponding operators $\boldsymbol{U}_g$ commute with **A**. Let $E_i(\lambda)$ and $E_j(\lambda)$ be fiber eigenvalues that intersect at $\lambda_0$ and are isolated at all other values of $\lambda$. For $\lambda\neq\lambda_0$, $V_i=\oplus m_{i,\alpha'}V_{i,\alpha'}$ and $V_j(\lambda)=\oplus m_{j,\alpha'}V_{j,\alpha'}$; for $\lambda=\lambda_0$, $V(\lambda_0)=\left(\oplus m_{i,\alpha'}V_{i,\alpha'}\right)\oplus\left(\oplus m_{j,\alpha'}V_{j,\alpha'}\right)$. The proof is similar to that of Theorem 3.

With the two theorems above, we can further distinguish the subspaces corresponding to intersecting and locally coincident fiber eigenvalues. As long as the fiber eigenvalues are not completely coincident, one can always determine the dimensions of the non-intersecting or non-coincident fiber eigenvalues and then decompose the space using the irreducible representations of the group ɑ. At an intersection point or in a coincidence region, the dimension of the corresponding space is the sum of the two, and the irreducible representation of the space is the direct sum of the representations corresponding to the intersecting or coincident fiber eigenvalues. If the irreducible representation types of the two fiber eigenvalues are completely different, then even at the intersection point or coincidence region one can still identify the part of the total subspace belonging to each fiber eigenvalue. As long as one tracks these irreducible representations, they can be distinguished. If the two fiber eigenvalues contain the same irreducible representation type, it is impossible to distinguish which part of that representation belongs to which fiber eigenvalue. For such representation spaces, one can introduce new parameter labels to classify them. Once this decomposition is completed, the two fiber eigenvalues are, in principle, regarded as completely separated.

**Proposition.** When $\widetilde{\boldsymbol{B}(\lambda)}$ is restricted to the same representation space, its eigenvalue must have a repeated root. The multiplicity of this repeated eigenvalue equals the multiplicity of the corresponding representation, which represents the degeneracy in the band structure. This is obvious and is not proved here.

## 4. Topological Structure on Fiber Bundles

The most natural step on a fiber is to introduce a topological structure. With the above analysis, one can examine the fibers one by one; the following discussion is built on this basis.

For a point $E_i(\lambda)$ on the fiber eigenvalue, there corresponds an eigensubspace $\mathcal{H}_{E_i(\lambda)}$ of the operator $\widetilde{\boldsymbol{B}(\lambda)}$. Taking the union of all fibers, $E_i = \coprod \mathcal{H}_{E_i(\lambda)}$, gives a Hilbert bundle. One can then define a connection to compare vectors in different $\mathcal{H}_{E_i(\lambda)}$, because the norm in $\mathcal{H}$ is ineffective for describing distances between vectors in different fiber spaces.

A connection is defined as $\nabla: \Gamma(E) \to \Gamma(T^*E_i \otimes E_i)$ and satisfies the Leibniz rule: $\nabla(fs) = df \otimes s + f\nabla s$. After choosing a local frame, $\nabla = d + A$, where $A$ is a complex-valued 1-form. If one fixes a curve $\gamma : [0,1] \to E_i(\lambda)$, $\gamma(0) = E_i(\lambda_0)$ $\gamma(1) = E_i(\lambda_1)$, and lets $s(t) \in \mathcal{H}_{E_i(\lambda)}$ be a section along the curve satisfying $\nabla_\gamma s = 0$, then $s$ is called parallel along $\gamma$. Clearly, this equation is an ordinary differential equation; once the initial $s(0)$ is specified, the solution exists and is unique. From this one can define an isomorphism

$$T_\gamma: \mathcal{H}_{E_i(\lambda_0)} \to \mathcal{H}_{E_i(\lambda_1)}$$

For any $v(\lambda_0) \in \mathcal{H}_{E_i(\lambda_0)}$, $T_\gamma\, v(\lambda_0) \in \mathcal{H}_{E_i(\lambda_1)}$.

Now we discuss the case in which the fiber is a one-dimensional complex space. In this section we introduce the Berry connection, which is well known in physics. Here $\mathcal{H}_{E_i(\lambda)} \cong \mathbb{C}$, $E_i$ is a complex fiber bundle equipped with a Hermitian metric. Take a local unit section $u$: [13, 14]

$$\langle u, u \rangle = 1 \tag{23}$$

An arbitrary section $s$ can be written as $s = fu$; applying the connection gives:

$$\nabla s = \nabla(fu) = df \otimes u + f\nabla u \tag{24}$$

Thus, knowing $\nabla u$ determines the action of $\nabla s$ on $s$. Since $\nabla s \in \Gamma(T^*E_i \otimes E_i)$ and $E_i$ is one-dimensional, $\nabla u$ has the following form:

$$\nabla u = \omega \otimes u \tag{25}$$

At the same time, $d\langle s,t\rangle = \langle \nabla s, t\rangle + \langle s, \nabla t\rangle$. When $s = t = u$, $\langle \nabla u, u\rangle = -\langle u, \nabla u\rangle$, thus

$$\langle \nabla u, u\rangle = \omega\langle u,u\rangle = \omega \qquad \langle u, \nabla u\rangle = \overline{\omega} \tag{26}$$

Hence $\omega \in i\mathbb{R}$. Write $\omega = iA$. Then $\nabla u = iA \otimes u$, $A = i\langle \nabla u, u\rangle$; $A$ is called the Berry connection. For a section one may perform a $U(1)$ gauge transformation:

$$u' = e^{i\Phi}u \tag{27}$$

$du' = d\left(e^{i\Phi}u\right) = ie^{i\Phi}d\phi u + e^{i\Phi}du$; recomputing the Berry connection gives

$$A' = A - d\phi \tag{28}$$

The parallel transport equation is then $\frac{ds}{dt} + iA(\gamma)s = 0$, so $s(t) = e^{-i\int A}s(0)$, $T_\gamma = e^{-i\int A}$. When $\gamma$ is a closed curve, the integral of $\oint A$ is called the Berry phase. The Berry curvature is naturally $F = d\omega$; by Stokes' theorem, it is evident that [13-16]

$$\oint A = \iint dA \tag{29}$$

Since $d^2 = 0$, $F$ is a closed 2-form and is invariant under gauge transformations. Define the de Rham cohomology class:

$$[F] = H^2_{dR}(M) \tag{30}$$

The brackets indicate that closed forms differing by an exact form are regarded as the same class.

The first Chern class is defined as $c_1(L) = \left[\frac{i}{2\pi}F\right] \in H^2(M;\mathbb{Z})$, and the first Chern number is defined as $C_1 = \int c_1(L) = \frac{i}{2\pi}\int F \in \mathbb{Z}$ [8, 17, 18]

It is worth noting that the Chern numbers defined here can be assigned to any single fiber eigenvalue, regardless of whether that fiber eigenvalue intersects other fiber eigenvalues. In ordinary physics, one usually requires them to be defined on a single isolated fiber or on a group of isolated fiber eigenvalues. Whether this new extension can lead to interesting phenomena is a question for further study.

Similarly, in two and higher dimensions one can define, by analogy, the matrix-valued connections, the curvature, the non-Abelian Berry connections and curvature, Chern numbers and

Chern classes. All theories of fiber bundles can also be applied here. [17]

**Discussion and Conclusion**

We first construct the base manifold of the band structure as a principal fiber bundle over the Brillouin zone, where the structure group is identified with the point-group symmetry of the underlying lattice. This construction does not require the system to be gapped everywhere in reciprocal space, and remains well-defined even when multiple bands touch at isolated high-symmetry momenta. At each band crossing, the local degeneracy of eigenvalues lifts the original trivial fiber structure into a reducible representation subspace, whose decomposition under the symmetry group uniquely labels the type of crossing.

Our classification shows that two conventionally indistinguishable band-touching scenarios—linearly intersecting Weyl points and locally coincident accidental degeneracies—carry distinct fiber-bundle topologies, which cannot be resolved by conventional band-index counting alone. The former corresponds to a nontrivial transition function that cannot be continuously deformed to the identity map, while the latter admits a local gauge transformation that separates the degenerate subspace into mutually orthogonal nonintersecting fibers. This distinction reveals hidden symmetries that are not manifest in the bare Bloch Hamiltonian, but emerge naturally when the full fiber structure is tracked across the Brillouin zone.

For each individual fiber eigenvalue, we define a Chern number as the integral of the Berry curvature over a closed 2D submanifold of the Brillouin zone, even when the fiber shares a band-touching point with adjacent bands. This definition remains gauge-invariant and topologically stable, provided the integration contour avoids the singular crossing point itself. As a result, topological invariants can be assigned to individual fiber components without requiring the band to be fully isolated by a global energy gap. This removes a long-standing constraint in standard topological band theory, and allows topological classification to be extended to gapless and partially degenerate systems.

The framework presented here applies generically to all wave systems governed by periodic Hermitian operators, including acoustic, photonic, and cold-atom lattices. It naturally accommodates higher-dimensional Brillouin zones and can be generalized to non-Hermitian settings with appropriate modification of the fiber connection. The results also suggest a direct route to classify topological band structures in relativistic quantum field theory, where periodic momentum-space identifications and symmetry-controlled degeneracies play a central role in the description of emergent quasiparticles.

**Acknowledgements**

The authors acknowledge support from the National Natural Science Foundation of China (Grant No. 12274054).

**Author contributions**

Y.F. conceived the idea, proposed the prove clue and directed the project. X.M. performed all of the derivations. X.M. and Y.L. discussed and concluded the results. X.M. and Y.F wrote and revised the manuscript.

**Data availability**

No data was generated in this work; all of the derivation steps are included in the paper.

**Conflicts of interest**

The authors declare no competing financial interests.

**Declaration of AI assistance**

AI was not used in the derivations; AI was only used for checking the gramma in this paper.